\documentclass[useAMS,usenatbib]{mn2e}
\usepackage[dvipdfmx]{graphicx}
\usepackage[dvipdfmx]{color}
\usepackage{aas_macros}

\newcommand{\nHII}{n_{\rm HII}}
\newcommand{\fesc}{f_{\rm esc}}

\newcommand{\Zsun}{\rm Z_{\odot}}
\newcommand{\Msun}{\rm M_{\odot}}
\newcommand{\Lsun}{\rm L_{\odot}}
\newcommand{\mum}{\rm \mu m}

\newcommand{\Loiii}{L_{{\rm O_{{\rm III}, 88}}}}

\newcommand{\OIII}{O{\sc iii}}
\newcommand{\HII}{H{\sc ii}~}
\newcommand{\CII}{C{\sc ii}}

\title[Far-infrared line emitters]
      {The distribution and physical properties of high-redshift [\OIII] emitters in a cosmological hydrodynamics simulation}

\author[K. Moriwaki et al.]
{Kana Moriwaki$^{1}$\thanks{E-mail: kana.moriwaki@utap.phys.s.u-tokyo.ac.jp},
Naoki Yoshida$^{1, 2, 3}$,
Ikkoh Shimizu$^{4}$,
Yuichi Harikane$^{1, 5}$, \and
Yuichi Matsuda$^{6, 7}$,
Hiroshi Matsuo$^{6, 7}$,
Takuya Hashimoto$^{6, 8}$,
Akio K. Inoue$^{8}$, \and
Yoichi Tamura$^{9}$,
Tohru Nagao$^{10}$.
\\
$^{1}$Department of Physics, The University of Tokyo, 
7-3-1 Hongo, Bunkyo, Tokyo 113-0033, Japan \\
$^{2}$Kavli Institute for the Physics and Mathematics of the Universe (WPI), 
UT Institutes for Advanced Study, \\
The University of Tokyo, 5-1-5 Kashiwanoha, Kashiwa, Chiba 277-8583, Japan \\
$^{3}$Research Center for the Early Universe, School of Science, The University of Tokyo, 
7-3-1 Hongo, Bunkyo, Tokyo 113-0033, Japan \\
$^{4}$Theoretical Astrophysics, Department of  Earth and Space Science, Osaka University,\\
1-1 Machikaneyama, Toyonaka, Osaka 560-0043, Japan \\
$^{5}$Institute for Cosmic Ray Research, The University of Tokyo, 5-1-5 Kashiwanoha, Kashiwa, Chiba 277-8582, Japan \\
$^{6}$National Astronomical Observatory of Japan, 2-21-1 Osawa, Mitaka, Tokyo 181-8588, Japan \\
$^{7}$Department of Astronomical Science, School of Physical Sciences, The Graduate University for Advanced Studies (SOKENDAI), \\
2-21-1 Osawa, Mitaka, Tokyo 181-8588, Japan \\
$^{8}$Department of Environmental Science and Technology, Faculty of Design Technology, \\
Osaka Sangyo University, 3-1-1, Nagaito, Daito, Osaka 574-8530, Japan \\
$^{9}$Division of Particle and Astrophysical Science, Graduate School of Science, Nagoya University, Furo-cho, Chikusa-ku, Nagoya 464-8602, Japan\\
$^{10}$Research Center for Space and Cosmic Evolution, Ehime University,
2-5 Bunkyo-cho, Matsuyama, Ehime 790-8577, Japan
}

\begin{document}

\date{}

\pagerange{\pageref{firstpage}--\pageref{lastpage}} \pubyear{0000}

\maketitle

\label{firstpage}

\begin{abstract}
 Recent observations with the Atacama Large Millimeter/submillimeter Array (ALMA)
 detected far-infrared emission lines such as the [\OIII] $88 \mum$ line from galaxies at $z \sim 7 - 9$.
 We use a cosmological simulation of galaxy formation to study the
 physical properties of [\OIII] $88 \mum$ emitters.
 In a comoving volume of $50 h^{-1}$ Mpc on a side, we locate 34 galaxies with stellar masses greater than $10^8\ \Msun$ 
 at $z = 9$, and more than 270 such galaxies at $z =  7$.
 We calculate the [\OIII] $88 \mum$ luminosities ($\Loiii$) by combining a physical model 
 of \HII  regions with emission line calculations using the photoionization code {\sc cloudy}.
 We show that the resulting $\Loiii$, for a given star formation rate, is slightly higher than 
predicted from the empirical relation for local galaxies, 
and is consistent with recent observations of galaxies at redshifts 7 - 9.
Bright [\OIII] emitters with $\Loiii > 10^8 ~\Lsun$ have
 star formation rates higher than $3 ~\Msun {\rm ~yr}^{-1}$, and the typical metallicity is $\sim 0.1 ~\Zsun$.
 The galaxies are hosted by dark matter halos with masses greater than $10^{11} ~\Msun$.
We propose to use the [\OIII] 5007$\rm \AA$ line, to be detected by James Webb Space
 Telescope (JWST), to study the properties of galaxies
  whose [\OIII] $88 \mum$ line emission have been already detected with ALMA.
\end{abstract}



\section{INTRODUCTION}
One of the major goals of the next generation space-borne and ground-based telescopes is to detect 
and characterize the first galaxies that were in place a few hundred million years after 
the Big Bang. 
There has been impressive progress in exploration of the redshift frontier of distant galaxies.
Hubble Space Telescope has detected a number of distant galaxies 
by dropout techniques 
\citep{Yan11, Ellis13, McLure13, Schenker13, Dunlop13, Robertson13, Ono13, Koekemoer13, Oesch13, Bouwens14, Finkelstein15, Bouwens15}.
Intrinsically faint galaxies have also been detected with the help of gravitational lensing magnification 
\citep{Zheng12, Oesch15, Ishigaki15}. 
Hydrogen Ly${\rm \alpha}$ line has been used primarily
to determine the spectroscopic redshifts of these galaxies.
Unfortunately, the Ly${\rm \alpha}$ line becomes increasingly weak at $z > 7$
through the epoch of reionization when the intergalactic medium (IGM) neutral fraction increases \citep[e.g.][]{Konno14}.

Rest-frame far-infrared lines are a promising probe of distant galaxies. The Atacama Large Millimeter/submillimeter 
Array (ALMA) is capable of detecting multiple far-infrared lines even from galaxies at $z > 7$.
Although initial attempts to detect the [\CII] 158$\mum$ line, one of the brightest emission lines
from star-forming galaxies, resulted in both success and non-detections
\citep{Walter12, Kanekar13, Ouchi13}, recent observations show
that the [\CII] 158$\mum$ line can be used to identify and to study the structure and kinematics
of distant galaxies \citep{Smit18}.

[\OIII] 88$\mum$ emission is another excellent probe of high-redshift star-forming
galaxies. From a theoretical point of view, the [\OIII] 88$\mum$ emission is easy to model 
because it originates from \HII regions, unlike the [\CII] 158$\mum$ emission which likely originates
from both ionized and neutral regions \citep{Nagamine06, Pallottini17, Lagache18}.
Interestingly, many nearby dwarf galaxies show stronger 
[\OIII] 88$\mum$ emission than [\CII] 158$\mum$ \citep{Madden12, Lebouteiller12, Cormier15}. 
Motivated by these observations, \citet{Inoue14a} proposed to use the [\OIII] 88$\mum$ line to determine the 
spectroscopic redshifts of galaxies at $z > 8$.
Recent observations using ALMA successfully discovered the [\OIII] line from galaxies beyond redshift 7
\citep{Inoue16, Carniani17, Laporte17} and even at $z = 9.11$ \citep{Hashimoto18a}. In combination with optical/infrared observations
and detection of the hydrogen Lyman-$\alpha$ line, the ALMA observations provide rich information
on the physical properties of the early galaxies that are thought to have driven cosmic reionization.

With the launch of James Webb Space Telescope (JWST) scheduled in 2021,
it is timely and important to study the formation and evolution of early emission line galaxies. 
To this end, we use a high-resolution cosmological simulation
of galaxy formation and study the statistics and physical properties of a population of 
[\OIII] emitters at $z > 7$. 
Throughout the present Letter, we adopt a ${\rm \Lambda CDM}$ cosmology with the matter
density $\Omega_{\rm M} = 0.3175$,
the cosmological constant $\Omega_{\rm \Lambda} = 0.6825$, the Hubble constant $h = 0.6711$
in units of $H_0 = 100 {\rm ~km ~s^{-1} ~Mpc^{-1}}$ and the baryon density $\Omega_{\rm B} = 0.04899$.
The matter density fluctuations are normalized by setting $\sigma_8 = 0.8344$ \citep{PlanckCollaboration14}.

\section{Methods}

We use outputs of a cosmological hydrodynamics simulation of \citet{Shimizu16}.
The simulation follows the formation of galaxies in a fully cosmological context by implementing
star formation,  the stellar feedback effects by radiation pressure and supernova explosions,
and multi-element chemical enrichment \citep{Okamoto08, Okamoto10, Okamoto14, Okamoto09}.
The feedback model parameters are chosen to reproduce the observed star-formation rate density
from $z=10$ to $z=0$, and 
the galaxy stellar mass functions, the stellar mass - halo mass relation, and
the relation between stellar mass and metallicity at $z = 4$ through to $z = 0$.
Our simulation reproduces a broad range of the observed features of high-redshift galaxy populations 
such as Lyman break galaxies, Lyman-${\rm \alpha}$ emitters, and sub-mm galaxies as well as
those of local galaxies \citep{Shimizu12, Shimizu14, Shimizu16, Okamoto14}. 
The initial conditions are configured with $2 \times 1280^3$ gas and dark matter particles in a cubic 
volume of comoving  $50 h^{-1}{\rm ~Mpc}$. The mass of a dark matter particle is $4.44\times 10^6 h^{-1} ~\Msun$ and the initial mass of 
a gas particle is $8.11 \times 10^5 h^{-1} ~\Msun$.
The softening length for the gravitational force is set to be $2 h^{-1}$ kpc in comoving unit. 
The details of the simulation are found in \citet{Shimizu16}.

Following the standard star-formation prescription, star particles are spawned in cold and dense gas clouds. 
The mass of a star particle is as small as $\sim 10^6 ~\Msun$, 
and thus galaxies with mass $\sim 10^9 ~\Msun$ are represented by more than $\sim 1000$ star particles
and gas particles. We study the internal structure of the galaxies on sub-kpc scales such as the relative distributions of young 
and old stars, and also the distribution of rest-frame optical/far-infrared line emitting regions. 

\begin{figure}
\begin{center}
  \includegraphics[width=7cm]{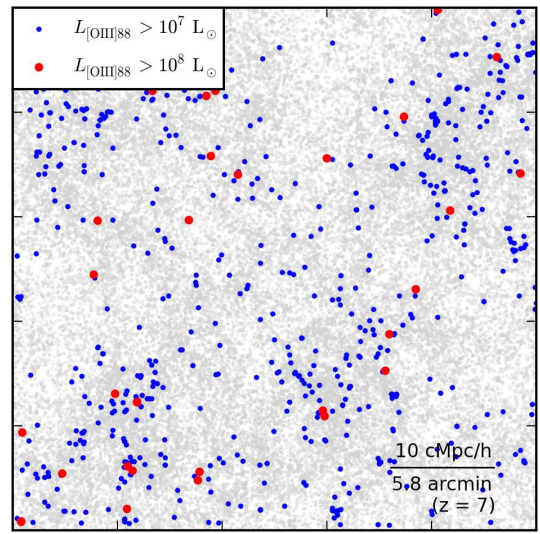}
\caption{
  We plot the projected distribution of [\OIII] emitters at $z = 7$ in a cubic volume of comoving
  $50 h^{-1}$ Mpc (= $29'$) on a side. The gray points are all the galaxies in this volume, whereas 
  the blue and red points represent galaxies with $\Loiii > 10^7 ~\Lsun$ and $> 10^8 ~\Lsun$, respectively.
  The [\OIII] 5007\AA \ line luminosities can be approximately estimated by multiplying $\Loiii$ by a factor of ten.
  }
\label{fig:clustering}
\end{center}
\end{figure}
	
\begin{table}
\caption{The parameters used to calculate the line luminosities with {\sc cloudy}. 
The metallicities are normalized by $\Zsun = 0.02$.
}
\centering
	\begin{tabular}{lc} \hline
	$\log_{10} (Z/\Zsun)$ & -2.2, -1.6, -0.6, -0.3, 0.1, 0.5  \\
	$\log_{10} U$ & -4.0, -3.9,  ..., -1.1, -1.0 \\
	$\log_{10} (\nHII/ {\rm cm^{-3}}) $ & 1.0, 2.0, 3.0 \\ \hline
	\end{tabular}
\label{number}
\label{parameter_table}
\end{table}

\section{Spectral energy distribution of galaxies}

We calculate the spectral energy distribution (SED) of each spawned star particle using the population synthesis code {\sc p\'{e}gase2} \citep{FiocRocca97}.
We obtain the total SED of a simulated galaxy by summing the contributions from all the star particles within the galaxy.
We adopt the Calzetti law of dust extinction \citep{Calzetti00}.
We calculate the escape probability $f_{\rm UV}$ of ultra-violet (UV) photons at 1500 ${\rm \AA}$ 
adopting a dust distribution model of \citet{XuBuat95} and \citet{Shimizu14}:
\begin{eqnarray}
	f_{\rm UV} = \frac{ 1 - \delta }{ 2 } ( 1 + e^{-\tau_{\rm d}} ) + \frac{\delta}{\tau_{\rm d}} ( 1 - e^{-\tau_{\rm d}} ),
\end{eqnarray}
where $\delta$ is a parameter whose value is from 0 to 1 and $\tau_{\rm d}$ is the UV optical depth.
The optical depth $\tau_{\rm d}$ is calculated in the same way as in \citet{Shimizu16}.
With $\delta = 0.95$ for attenuation of continuum, the rest-frame UV luminosity function matches well
to the observed ones at $z=6-10$. 

The nebular emission line luminosities are calculated in the following manner.
We generate a library of emission lines using {\sc cloudy}
\footnote{
We configure plane-parallel geometry, and stop the calculations 
when the fraction of the electron number density and the hydrogen number density becomes lower than $10^{-3}$.
To generate the source SEDs, we adopt constant star formation with Salpeter initial mass function
and use {\sc bc03} \citep{BruzualCharlot03}. 
Note that the difference beween {\sc bc03} and {\sc p\'{e}gase2} is very small.
}
\citep{Ferland13} as in \citet{Inoue11} and \citet{Inoue14a}.
The library covers a wide range of gas metallicity $Z$, ionization parameter $U$ and gas density $\nHII$
as given in Table \ref{parameter_table}.
We use the local oxygen abundance $y_{\rm O}$ to derive the corresponding metallicity 
as $Z = \Zsun$ $ \times\ y_{\rm O} / y_{{\rm O}, \odot}$, where $y_{{\rm O}, \odot}$ is the solar abundance of oxygen.

For the current and future use, the library lists the line luminosity, $L_{\rm line}$, such that it is normalized
by the ${\rm H\beta}$ luminosity with the case-B approximation (Dopita \& Sutherland 2003), $L_{\rm H\beta}^{\rm caseB}$, as
\begin{eqnarray}
	L_{\rm line} = ( 1 - \fesc )C_{\rm line}(Z, U, \nHII) L_{\rm H\beta}^{\rm caseB},
	\label{eq:em}
\end{eqnarray}
where $\fesc$ is the Lyman continuum escape fraction and $C_{\rm line}$ is the line luminosity ratio.
We assume $\fesc = 0.1$ in the present Letter. The choice of a constant escape fraction may be unrealistic,
since it is intrinsically a time-dependent quantity and depends also on a variety of properties such as halo mass and gas clumping.
Recent high-resolution radiation hydrodynamics simulations suggest time-averaged escape fractions of
several to ten percent for high-redshift galaxies \citep{Trebitsch17}, which motivate our choice here.
Since the line luminosities are proportional to $(1- \fesc)$,  
other cases with higher/lower $\fesc$ can be inferred by scaling the line luminosities
correspondingly.

Individual \HII regions are not resolved in our simulation, and thus we resort to 
a simple physical model of the inter-stellar medium (ISM) structure to calculate the emissivities
of the lines originating from \HII regions.
We characterize the ISM by the local gas density $n$ and metallicity $Z$, and also by a volume-averaged ionization parameter
\begin{eqnarray}
	U = \frac{3\alpha_{\rm B}^{2/3}}{4c} \Big( \frac{3 Q \nHII}{4\pi}\Big)^{1/3},
\end{eqnarray}
where $Q$ is the production rate of ionizing photons from each star particle, and $\alpha_{\rm B}$
is the case-B hydrogen recombination coefficient. 
Essentially, we assume a constant gas density $\nHII$ in a spherical 
\HII region surrounding a star particle \citep[e.g.][]{Panuzzo03, Hirschmann17}.
We assume that the \HII region density is given by $\nHII = K n$,
where $n$ is the gas particle density in the vicinity of the star particle,
so that compact galaxies tend to have large $\nHII$.
With $K=5$ as our fiducial value,
$\nHII$ ranges approximately from 30 to 300 ${\rm cm}^{-3}$,
which is the typical range of the \HII region density \citep[e.g.][]{Osterbrock06, Brinchmann08}.
We note that different methods, as well as different numerical resolutions, may slightly change the estimated [OIII] luminosities.

\begin{figure}
\begin{center}
  \includegraphics[width=8cm]{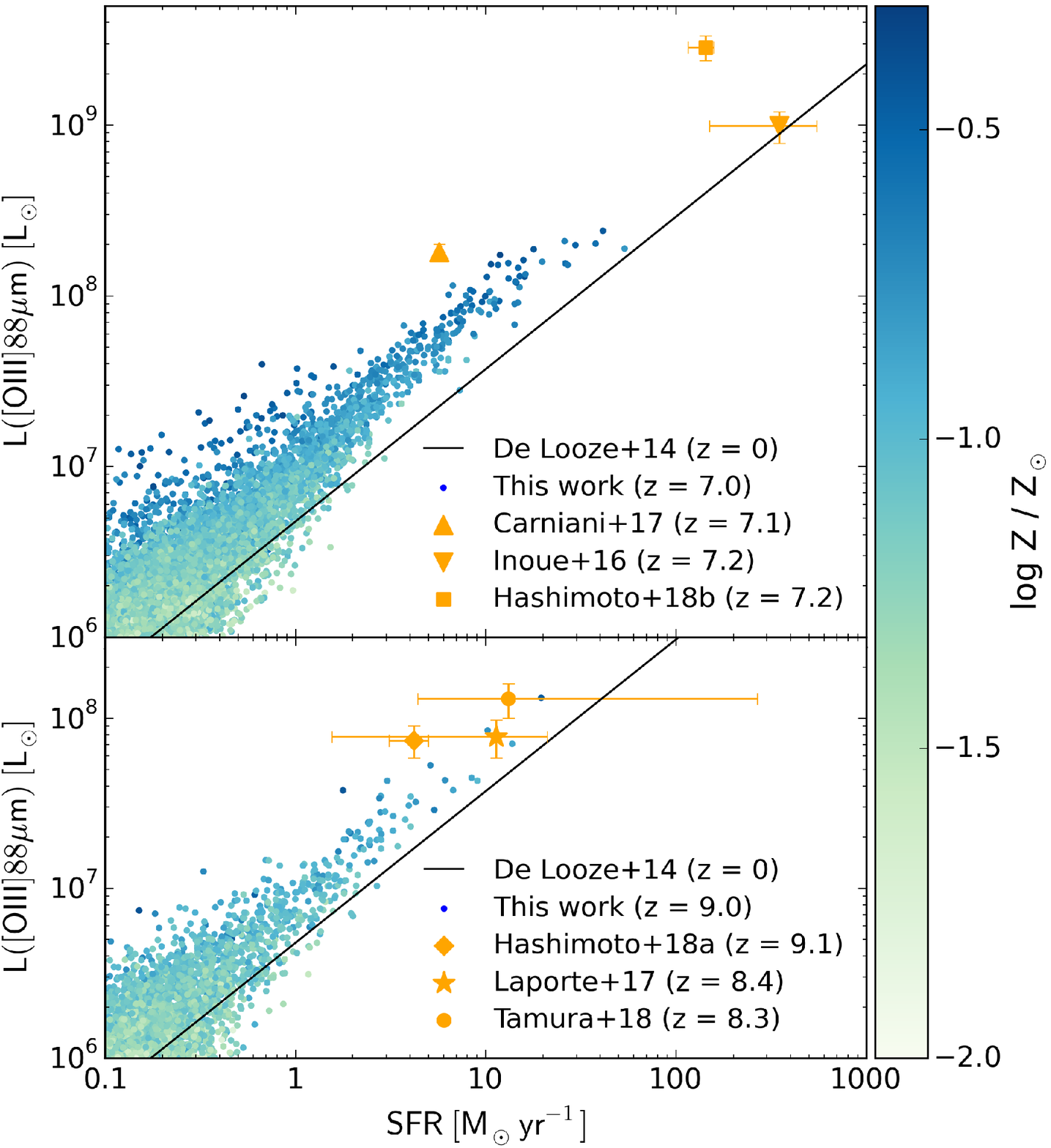}
  \caption{The [\OIII] line luminosity against star formation rate of our galaxy samples at $z = 7$
    and $z=9$. We indicate the gas metallicity by the color of each point.
    We compare them with the observed galaxies at $z = 7 - 9$ 
    \citep{Inoue16, Carniani17, Laporte17, Hashimoto18a, Hashimoto18b, Tamura18}.
    The black line shows the [\OIII]-SFR relation derived by \citet{DeLooze14} for all kinds of local galaxies. 
    Note that the galaxy B14-65666 of \citet{Hashimoto18b} at $z = 7.15$ appears to consist of two clumps.}
  \label{fig:oiii-sfr}
\end{center}
\end{figure}

\section{Results}
Figure \ref{fig:clustering} shows the spatial distribution of the [\OIII] emitters identified at $z=7$.
The blue and red points represent galaxies with $\Loiii > 10^7 ~\Lsun$ and $> 10^8 ~\Lsun$, respectively.
The [\OIII] emitters are strongly clustered, and trace the large-scale structure of $\sim 10$ Mpc scales. 
We calculate the two-point correlation function of the bright [OIII] emitters.
The clustering bias of galaxies with $\Loiii > 10^7 ~\Lsun$ is 4.0 at a separation of $3 h^{-1}$ comoving mega-parsec.

The early large-scale structure and its evolution can be probed by future observations 
with JWST NIRCam targeting rest-frame optical emission lines such as [\OIII] 5007\AA, 
as we propose in the Discussion section,
or by utilizing the intensity mapping technique at sub-millimeter wavelengths \citep{Visbal10}.
Our simulation predicts that the total [\OIII] 88$\mum$ luminosity at $z = 7$ in a cubic volume 
of comoving $50h^{-1}$ Mpc is $2.2\times 10^{10} ~\Lsun$, giving a significant contribution to the 
global sub-millimeter line intensity.  

In Figure \ref{fig:oiii-sfr}, we plot $\Loiii$ against star-formation rate (SFR) for our emission line galaxy samples. 
For reference, we also show the $\Loiii$-SFR relation for all kinds of local galaxies derived by \citet{DeLooze14}.
Our model predicts slightly larger values of $\Loiii$ than the local relation; early galaxies tend to be
compact and have large ionization parameters.  
Several galaxies in the simulated volume have $\Loiii$ and SFRs comparable to 
the intrinsic $\Loiii$ and SFRs of the galaxies recently identified 
at $z \sim 7 - 9$ \citep{Carniani17, Laporte17, Hashimoto18a}. 
The bright [\OIII] emitters have metallicities of $\sim 0.1\ \Zsun$, consistent with
the estimate of, e.g., \citet{Inoue16}. 
The [\OIII] emitters with $\Loiii > 10^8 ~\Lsun$ at $z=7$ are hosted by dark matter halos
with mass greater than $10^{11} ~\Msun$. 

\begin{figure}
\begin{center}
  \includegraphics[width=8.5cm]{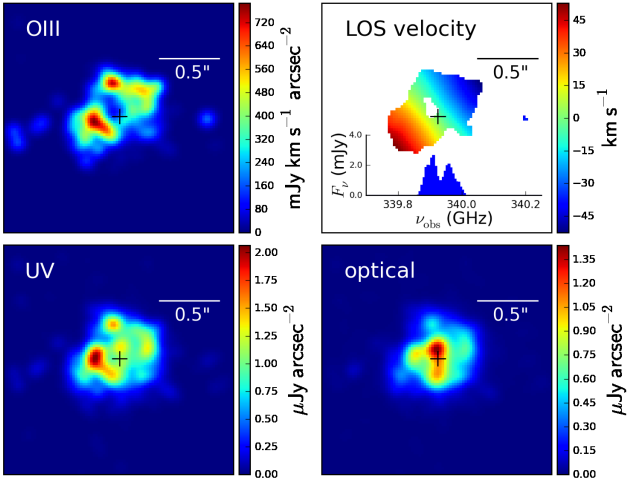}
\caption{
The structure of a galaxy at $z = 9$ with $M_* = 6.3\times 10^8 ~\Msun$, $M_{\rm halo} = 8.2\times 10^{10} ~\Msun$, 
SFR = 4.9 $\Msun/{\rm yr}$,  and $\Loiii = 5.4\times 10^7 ~\Lsun$.
We plot the [\OIII] 88$\mum$ flux, the [\OIII] luminosity weighted line-of-sight gas velocity, 
the rest-frame UV continuum (1600 \AA) flux and the rest-frame optical continuum (5000 \AA) flux.
The cross indicates the mass center of the stars.
For this figure, we assume that the galaxy is gravitationally lensed with $\mu=10$,
to be compared with, e.g., MACS1149-JD1 \citep{Hashimoto18a}.
The inset of the upper right panel shows the spectrum to be observed with ALMA.
}
\label{fig:colormap}
\end{center}
\end{figure}

Some bright [\OIII] emitters have extended structure of $\sim$ 1 physical kilo-persec.
We find that the [\OIII] 88$\mum$ emitting regions are localized 
and sometimes displaced from the bulk stellar distribution.
This is because the stellar distribution traces the regions where prior star-formation 
activities have already consumed the gas, while
on-going star-formation in other places powers the instantaneous [\OIII] emission.
Figure \ref{fig:colormap} shows the structure of one of the brightest galaxies at $z=9$. 
Motivated by the intriguing observation of a gravitationally lensed galaxy 
MACS1149-JD1 \citep{Hashimoto18a}, we generate Figure \ref{fig:colormap} assuming an isotropic
lensing magnification of $\mu=10$. With the help of lensing, the effective physical resolution 
increases by a factor of $\sim 3$.

Both of the [\OIII] 88$\mum$ line emission and the UV continuum trace 
young stellar populations with ages shorter than several million years.
Though the relative displacement between the two distributions is often found for our bright galaxy samples,
the flux peaks roughly coincide with each other, consistent with the observed galaxies
at $z \sim 7 - 9$ \citep{Inoue16, Hashimoto18a}.
We note that there can also be spatial offsets between the [\OIII] 88$\mum$ and the UV 
emission caused by an inhomogeneous dust distribution \citep{Katz17}.

\section{Discussion}

\begin{figure}
\begin{center}
  \includegraphics[width=8cm]{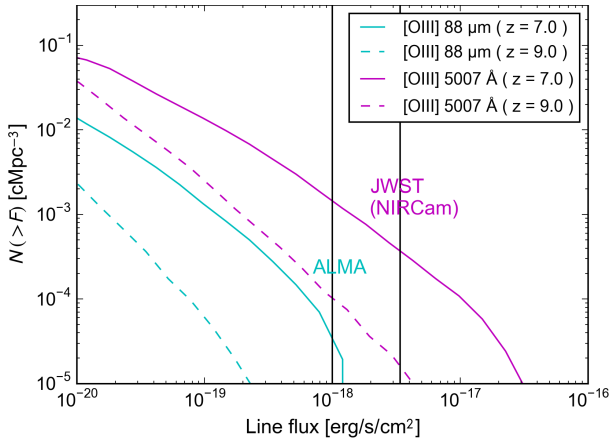}
\caption{The cumulative line luminosity function for [\OIII] 88$\mum$ (cyan lines) 
  and [\OIII] 5007
  \AA (magenta lines). 
  The solid lines and the dotted lines show the data of $z = 7$ and $z = 9$ respectively.
  Vertical lines represent typical detection limits of ALMA and JWST NIRCam grism mode
  assuming $10^4$ second integration and the signal-to-noise ratio of 5.
}
\label{fig:lf_oiii}
\end{center}
\end{figure}

Conventional optical line diagnostics can be used to study the physical properties of the early star-forming galaxies.
We calculate the [\OIII] 5007$\rm \AA$ luminosity function and compare it with the luminosity function 
of [\OIII] 88$\mum$ (Figure \ref{fig:lf_oiii}). The vertical lines in the figure indicate typical detection limits of ALMA and JWST.
As expected, the [\OIII] 5007$\rm \AA$ luminosity is roughly an order of magnitude larger than that of [\OIII] 88$\mum$.
Clearly, JWST can detect [\OIII] 5007$\rm \AA$ line emission from the galaxies whose [\OIII] 88$\mum$ lines 
(will) have been detected with ALMA. 

The superb spectral and angular resolution of JWST NIRSpec IFU spectroscopy will enable us
to study the structure and the kinematics of galaxies at $z>7$.
Gravitational lensing greatly helps not only by increasing the apparent flux of distant
galaxies, but also by increasing the effective 'physical' spatial resolution. 
The sub-arcsecond resolutions of ALMA and JWST can be fully exploited to study 
the fine structure of lensed galaxies even at sub-kiloparsec scales.  
Signatures of the galaxy's rotation can be also seen
as the line-of-sight velocity gradient of $\Delta v \sim 100 {\rm ~km ~s^{-1}}$ 
(see the upper-right panel of Figure \ref{fig:colormap}).

Future observations with ALMA and JWST hold promise to understand the formation and evolution
of the first galaxies. Combining with fully resolved observations of other rest-frame optical and far-infrared 
emission lines originating from hydrogen, carbon and oxygen in a variety of phases in the ISM, 
and also with dust continuum emission, 
we will be able to study the chemical 
evolution and the ISM structure and kinematics of the galaxies a few to several hundred million years after the
Big Bang. It is important to explore theoretically pan-chromatic approaches to elucidate the nature of the first galaxies.

 The large-scale distribution of the [\OIII] emitters can be probed by using JWST NIRCam, either by using
 its narrow band filters or with its grism spectroscopy mode. For instance, the line sensitivity for the NIRCam 
 grism module at the F444W band is $\sim 3.4\times 10^{-18} {\rm ~erg ~s^{-1} ~cm^{-2}}$ for S/N = 5 
with a $10^4$ second exposure. It is expected from Figure \ref{fig:clustering} and \ref{fig:lf_oiii} that, within the field-of-view of $2 \times 2.2' \times 2.2'$, 
there will be several [\OIII] 5007\AA\ emitters on average detected in the redshift range of $6.8 < z < 9.0$. 
Multiple pointings or mosaic observations will map the
three-dimensional distribution of [\OIII] emitters in the redshift range.

Systematic surveys with JWST NIRCam narrow-bands can also probe the large-scale structure at $z>8$.
 If we perform a narrow-band survey using the F466N filter with the same exposure, we can detect, with 
 S/N = 5, $z=8.3$ [\OIII] emitters with line fluxes above 
 $\sim 6.6\times 10^{-19} {\rm ~erg ~s^{-1} ~cm^{-2}}$. This limit roughly corresponds to 
 detecting the galaxies shown by blue points in Figure \ref{fig:clustering}. 
 In such narrow-band surveys, foreground H$\alpha$ emitters are
 severe contaminants, or can actually be major targets at the corresponding wavelength \citep{Silva18}, but
 one can use $J$-dropout technique to select [\OIII] 5007$\rm \AA$ emitters. 
 We have found that the [\OIII] emitters at $z=8.3$ in our simulation are 
 separated typically by $Y_{105} - J_{125} > 1.0-1.5$ from H$\alpha$ emitters at  $z=6.2$. 
 It is possible to select preferentially the [\OIII] emitters if we choose and survey a region where deep broad-band data
 are available. 
 
 Future space missions such as Spectrophotometer for the History of the Universe, Epoch of Reionization, and
 Ice Explorer (SPHEREx) and Cosmic Dawn Intensity Mapper (CDIM) utilize the infrared intensity mapping technique to
 probe the large-scale galaxy distribution at high-redshifts \citep{Cooray16, Dore16}. 
 Redshifted [\OIII] 5007\AA\ emission from early star-forming galaxies such as those studied in the present Letter
 is an excellent target, and thus serve as a valuable cosmological probe.
 
\section*{acknowledgments}
The authors thank the anonymous referee for providing us with many insightful comments.
This research was supported by the Munich Institute for Astro- and Particle Physics (MIAPP) of 
the DFG cluster of excellence "Origin and Structure of the Universe".
K.M. has been supported by Advanced Leading Graduate Course for Photon Science
(ALPS) of the University of Tokyo.
I.S., Y.H., Y.M., A.K.I. and Y.T. acknowledge JSPS KAKENHI grants 26247022, 17H01111 (I.S.), 16J03329 (Y.H.), 17H04831, 17KK0098 (Y.M.), 17H01114 (A.K.I) and 17H06130 (Y.T).
T.H., A.K.I and Y.T. acknowledge the NAOJ ALMA Scientific Research Grand number 2016-01A (T.H., A.K.I) and 2018-09B (Y.T.).


\bibliography{bibtex_library}

\end{document}